\documentstyle[preprint,revtex]{aps}
\tightenlines
\begin{document}

\draft

\preprint{SLAC--PUB--5915}
\medskip
\preprint{WIS--92/70/Sep--PH}
\medskip
\preprint{September 1992}
\medskip
\preprint{T/E}

\begin{title}
QCD Sum Rule Analysis of the Subleading Isgur-Wise\\
Form Factor $\chi_2(v\cdot v')$
\thanks{Work supported by
the Department of Energy, contract DE-AC03-76SF00515.}
\end{title}

\author{Matthias Neubert}
\begin{instit}
Stanford Linear Accelerator Center\\
Stanford University, Stanford, California 94309
\end{instit}
\author{Zoltan Ligeti and Yosef Nir}
\begin{instit}
Weizmann Institute of Science\\
Physics Department, Rehovot 76100, Israel
\end{instit}

%\receipt{September 1992}

\begin{abstract}
We present a QCD sum rule calculation of the spin-symmetry violating
universal function $\chi_2(v\cdot v')$, which appears at order $1/m_Q$
in the heavy quark expansion of meson form factors. This function
vanishes in the standard approximation, where radiative effects are
neglected. For the first time, the complete set of diagrams arising at
order $\alpha_s$ is evaluated. In particular, we find $\chi_2(1) =
-(3.8\pm 0.4)\%$ at zero recoil, indicating that $1/m_Q$ corrections
induced by the chromo-magnetic moment operator are small.
\end{abstract}
%\pacs{11.50.Li, 13.20.-v, 11.30.Hv, 11.30.Ly}
\centerline{(submitted to Physics Letters B)}
\newpage

\narrowtext

\section{Introduction}

Recently, the discovery of a spin-flavor symmetry that QCD reveals for
heavy quarks has increased the prospects both for a reliable
determination of some of the weak mixing angles and a study of
nonperturbative QCD in semileptonic decays of heavy mesons
\cite{Volo,Isgu}. These symmetries are responsible for restrictive
relations among weak decay amplitudes and reduce the number of
independent form factors. The heavy quark effective theory provides a
convenient framework in which to analyze such processes
\cite{Geor,Mann,Falk,FGL}. It allows a systematic expansion of decay
amplitudes in powers of $1/m_Q$. The coefficients in this expansion are
universal functions of the kinematic variable $y=v\cdot v'$, where $v$
and $v'$ denote the velocities of the initial and final mesons. They
are independent of the masses and spins of the heavy quarks, but
characterize the properties of the light degrees of freedom. At leading
order, only a single function $\xi(y)$ appears. The conservation of the
vector current implies that this celebrated Isgur-Wise form factor is
normalized at zero recoil, allowing model-independent predictions
unaffected by hadronic uncertainties.

Already at order $1/m_Q$, however, one encounters a set of four
additional universal functions $\xi_3(y)$ and $\chi_i(y)$ for $i=1,2,3$,
as well as a parameter $\bar\Lambda$, which denotes the mass difference
between a heavy meson and the heavy quark that it contains \cite{Luke}.
These functions break the spin-flavor symmetries and induce corrections
to the relations valid in the infinite quark mass limit. At zero
recoil, two of them can be shown to vanish, $\chi_1(1)=\chi_3(1)=0$
\cite{Luke}, but $\xi_3(1)$ and $\chi_2(1)$ are nonzero and lead to
observable effects. For instance, the leading correction to the ratio
of the axial form factors describing $B\to D^*\ell\,\nu$ decays is
given by \cite{Sublea}
\begin{equation}\label{ratio}
   {4 m_B m_{D^*}\over(m_B + m_{D^*}\!)^2}\,
   {A_2(q_{\rm max}^2)\over A_1(q_{\rm max}^2)} = 1 -
   {\bar\Lambda\over 2 m_c}\,\Big[ (1+3 r)\,\xi_3(1) +
   4(1-r)\,\chi_2(1) \Big] ,
\end{equation}
where $r=m_c/m_b$. An understanding of the universal form factors
is at the heart of nonperturbative QCD, but is necessary for any
phenomenological application of the heavy quark expansion. QCD sum
rules are particularly suited for this purpose and have recently been
employed to calculate the Isgur-Wise form factor
\cite{Buch,Rady,SR,Blok}, the mass parameter $\bar\Lambda$
\cite{BBBD,SR2}, and the universal functions that appear at order
$1/m_Q$ in the heavy quark expansion \cite{Sublea}. For these
subleading form factors an interesting result was obtained. Within
the approximations usually adopted in QCD sum rules, one finds the
parameter-free predictions \cite{Sublea}
\begin{equation}\label{predic}
   \xi_3(1) = \case{1}/{D-1} = \case{1}/{3} ,\qquad
   \chi_2(y) = 0 ,
\end{equation}
where $D$ is the space-time dimension. Corrections to these relations
could only come from radiative effects or from condensates of rather
high dimension, both of which are expected to be small and are usually
neglected. Provided those corrections are indeed negligible, the
predictions (\ref{predic}) allow for an essentially model-independent
description of the $1/m_Q$ corrections at zero recoil. For instance,
using $r\simeq 1/3$ and $\bar\Lambda\simeq 0.5$ GeV, one finds $\simeq
0.9$ for the right-hand side of (\ref{ratio}).

In this paper we investigate the corrections to the second relation in
(\ref{predic}), which arise if one goes beyond the standard
approximations by including diagrams of order $\alpha_s$ in the sum
rule for $\chi_2(y)$. In particular, we calculate for the first time
the two-loop radiative corrections to the perturbative triangle diagram.

\section{Derivation of the Sum Rule}

The heavy quark effective theory provides an expansion of hadronic
matrix elements in powers of $1/m_Q$. Its construction is based on the
observation that, in the limit $m_Q\gg\Lambda_{\rm QCD}$, the velocity
$v$ of a heavy quark $Q$ is conserved with respect to soft processes
\cite{Geor}. In this limit it is possible to remove the $m_Q$-dependent
piece of the heavy quark momentum $P_Q$ by the field redefinition
\begin{equation}
   h_Q(v,x) = e^{i m_Q v\cdot x}\,P_+\,Q(x)
\end{equation}
where $P_+=\case{1}/{2}(1+\rlap/v)$ is a positive-energy projection
operator. For simplicity we will abbreviate $h=h_Q(v,x)$. The new field
$h$ carries the residual ``off-shell'' momentum $k=P_Q-m_Q v$, which is
of order $\Lambda_{\rm QCD}$. The effective Lagrangian for the strong
interactions of the heavy quark becomes \cite{Geor,Mann,AMM}
\begin{equation}\label{Leff}
   {\cal{L}}_{\rm HQET} = \bar h\,i v\!\cdot\!D\,h + {1\over 2 m_Q}
   \bigg[ \bar h\,(i D)^2 h + Z\,{g_s\over 2}\,
   \bar h\,\sigma_{\alpha\beta} G^{\alpha\beta} h \bigg] + \ldots ,
\end{equation}
where $Z$ is a renormalization factor, $D=\partial - i g_s t_a A_a$ is
the gauge-covariant derivative, and $G$ is the gluon field strength.

A similar expansion can be performed for a generic heavy quark current
$\bar Q'\,\Gamma\,Q$, with $\Gamma$ being a $4\times 4$ Dirac matrix
carrying any number of Lorentz indices ($\Gamma=\gamma_\mu(1-\gamma_5)$
for the weak flavor-changing current). At tree-level, it simply reads
\begin{equation}\label{Jexp}
   \bar Q'\,\Gamma\,Q \to C\,\bar h'\,\Gamma\,h + \ldots ,
\end{equation}
where $h'=h_{Q'}(v')$, and $C$ is again a renormalization factor
\cite{Falk,QCD}.

A convenient way to evaluate hadronic matrix elements in the effective
theory is by using the covariant trace formalism introduced in
Refs.~\cite{Falk,Bjor}. Matrix elements of the leading current operator
in (\ref{Jexp}) can be parameterized as ($y=v\cdot v'$)
\begin{equation}
   \langle M'(v') |\,\bar h'\,\Gamma\,h\,| M(v) \rangle
   = - \xi(y)\, {\rm tr}\big\{\,
   \overline{\cal{M}}'(v')\,\Gamma\,{\cal{M}}(v) \big\} ,
\end{equation}
where $\xi(y)$ is the Isgur-Wise function, and
\begin{equation}
   {\cal{M}}(v) = \sqrt{m_M}\,P_+\, \cases{
    -\gamma_5 &; pseudoscalar meson \cr
    \rlap/\epsilon &; vector meson \cr}
\end{equation}
is a spin wave function associated with the pseudoscalar or vector
meson $M$, which has the right transformation properties under Lorentz
boosts and heavy quark spin rotations. At subleading order in the
$1/m_Q$ expansion, matrix elements receive contributions from
higher-dimension operators in the effective Lagrangian (\ref{Leff}) and
in the effective current (\ref{Jexp}). A particular type of correction
comes from an insertion of the chromo-magnetic operator contained in
${\cal{L}}_{\rm HQET}$. The corresponding matrix elements of
\begin{equation}
   {\cal{O}}(x) = i\!\int\!{\rm d}y\,T\Big\{
   \big[\, \bar h'\,\Gamma\,h \,\big]_x,
   \big[\, \case{g_s}/{2}\,\bar h\,\sigma_{\alpha\beta}
   G^{\alpha\beta} h \,\big]_y \Big\}
\end{equation}
can be parameterized by a tensor form factor,
\begin{equation}\label{chidef}
   \langle M'(v') |\,{\cal{O}}\,| M(v) \rangle = - \bar\Lambda\,
   {\rm tr}\Big\{ \chi^{\alpha\beta}(v,v')\,\overline{\cal{M}}'(v')\,
   \Gamma\,P_+\,i\sigma_{\alpha\beta} {\cal{M}}(v) \Big\} .
\end{equation}
Because $v^\alpha P_+ \sigma_{\alpha\beta} {\cal{M}}(v) = 0$, the
most general decomposition of $\chi^{\alpha\beta}$ consistent with
Lorentz invariance and the heavy quark symmetries involves two real,
scalar functions $\chi_2$ and $\chi_3$ defined by \cite{Luke}
\begin{equation}
   \chi^{\alpha\beta}(v,v')
   = (v'^\alpha\gamma^\beta - v'^\beta\gamma^\alpha)\,\chi_2(y)
   - 2 i\sigma^{\alpha\beta}\,\chi_3(y) .
\end{equation}
Since we have factored out $\bar\Lambda$ in (\ref{chidef}), the
functions $\chi_i(y)$ are dimensionless.

The QCD sum rule analysis for these subleading form factors proceeds
along the same lines as that for the Isgur-Wise function. The procedure
is outlined in detail in Refs.~\cite{Sublea,SR,SR2}, and we shall adopt
the same notations here. We are interested in the analytic properties
of the three-current correlator
\begin{eqnarray}\label{correl}
   \Xi &=& \int\!{\rm d}x\,{\rm d}y\,e^{i(k'\cdot x - k\cdot y)}\,
    \langle 0\,|\,T\Big\{
    \big[\, \bar q\,\overline{\Gamma}_{M'} h' \,\big]_x,
    {\cal{O}}(0),
    \big[\, \bar h\,\Gamma_M\,q \,\big]_y \Big\} |\,0\,\rangle
    \nonumber\\
   &=& \Xi_2(\omega,\omega',y)\,{\rm tr}\Big\{
    (v'^\alpha\gamma^\beta - v'^\beta\gamma^\alpha)\,
    \overline{\Gamma}_{M'} P_+' \Gamma\,P_+\,i\sigma_{\alpha\beta}
    P_+ \Gamma_M \Big\} \nonumber\\
   &+& \Xi_3(\omega,\omega',y)\,{\rm tr}\Big\{
    2 \sigma^{\alpha\beta}\,\overline{\Gamma}_{M'} P_+' \Gamma\,P_+
    \sigma_{\alpha\beta} P_+ \Gamma_M \Big\} ,
\end{eqnarray}
where $P_+'=\case{1}/{2}(1+\rlap/v')$. Depending on the choice
$\Gamma_M = -\gamma_5$ or $\Gamma_M = \gamma_\mu - v_\mu$, the
heavy-light currents interpolate pseudoscalar or vector mesons,
respectively. In the effective theory, the coefficients $\Xi_i$ are
analytic functions in the ``off-shell energies'' $\omega=2 v\cdot k$
and $\omega'=2 v'\cdot k'$, with discontinuities for positive values of
these variables. They receive a double-pole contribution from the
ground-state mesons $M$ and $M'$ associated with the heavy-light
currents. Using the fact that the external momenta are $P=m_Q v+k$ and
$P'=m_{Q'} v'+k'$, one finds that the pole positions $P^2=m_M^2$ and
$P'^2=m_{M'}^2$ correspond to $\omega=\omega'= 2\bar\Lambda$, where
$\bar\Lambda = m_M-m_Q = m_{M'}-m_{Q'}$ \cite{Luke}. The residues are
proportional to the universal functions $\chi_2(y)$ and $\chi_3(y)$,
respectively. It follows that \cite{Sublea}
\begin{equation}\label{pole}
   \Xi_i^{\rm pole}(\omega,\omega',y) = {\chi_i(y)\,F^2\,\bar\Lambda
   \over (\omega-2\bar\Lambda+i\epsilon)\,
   (\omega'-2\bar\Lambda+i\epsilon)} ~;~~i=2,3 ,
\end{equation}
where $F$ is defined by $\langle\,0\,|\,\bar q\,\Gamma\,h\,| M\rangle
= \case{i}/{2} F\,{\rm tr}\{\Gamma\,{\cal{M}}\}$. It is the analog of
the meson decay constant in the effective theory \cite{SR}.

For large negative values of $\omega$ and $\omega'$, the correlator can
be calculated in perturbation theory using the Feynman rules of the
heavy quark effective theory \cite{Falk}. The idea of QCD sum rules is
that, at the transition from the perturbative to the nonperturbative
region, nonperturbative effects can be accounted for by including the
leading power corrections in the operator product expansion of the
three-point function. They are proportional to vacuum expectation
values of local quark-gluon operators, the so-called condensates
\cite{SVZ}. One then writes the theoretical expression for $\Xi_i$ in
terms of a double dispersion integral,
\begin{equation}
   \Xi_i^{\rm th}(\omega,\omega',y) = \int\!{\rm d}\nu\,{\rm d}\nu'\,
   {\rho_i^{\rm th}(\nu,\nu',y)\over(\nu-\omega-i\epsilon)\,
   (\nu'-\omega'-i\epsilon)} + {\rm subtractions} ,
\end{equation}
and performs a Borel transformation
\begin{equation}
   {1\over\tau}\,\widehat{B}_\tau^{(\omega)}
   = \!\!\lim_{\matrix{ n\to\infty \cr
                        -\omega\to\infty \cr}}\!\!
   {\omega^n\over\Gamma(n)}\,
   \bigg(-{{\rm d}\over{\rm d}\omega}\bigg)^n
   ~;~~ \tau = {-\omega\over n}~{\rm fixed}
\end{equation}
with respect to $\omega$ and $\omega'$. This yields an exponential
damping factor in the dispersion integral and eliminates possible
subtraction terms. Because of the flavor symmetry it is natural to set
the Borel parameters equal, $\tau=\tau'=2 T$. It is then convenient to
introduce new variables $\omega_\pm=\case{1}/{2}(\nu\pm\nu')$ to obtain
\begin{equation}
   \widehat{\Xi}_i^{\rm th} = \widehat{B}_{2T}^{(\omega)}
   \widehat{B}_{2T}^{(\omega')}\,\Xi_i^{\rm th}
   = \int\!{\rm d}\omega_+\,e^{-\omega_+/T}
   \int\!{\rm d}\omega_-\,2\rho_i^{\rm th}
   \big(\omega_+\!+\!\omega_-,\omega_+\!-\!\omega_-,y\big) .
\end{equation}
Following Refs.~\cite{SR,Blok} we perform the integral over $\omega_-$
and employ quark-hadron duality to equate the integral over $\omega_+$
up to a threshold $\omega_0$ to the Borel transform of the pole
contribution in (\ref{pole}). This gives the QCD sum rules
\begin{equation}\label{sumrul}
   \chi_i(y)\,F^2\,\bar\Lambda\,e^{-2\bar\Lambda/T}
   = \int\limits_0^{\omega_0}\!{\rm d}\omega_+\,e^{-\omega_+/T}\,
   \widetilde{\rho}_i^{\rm\,th}(\omega_+,y) ~;~~i=2,3 .
\end{equation}
The spectral densities $\widetilde{\rho}_i^{\rm\,th}$ arise after
integration of the double spectral densities over $\omega_-$.

In this paper we concentrate on the sum rule for $\chi_2(y)$. It has
been shown in Ref.~\cite{Sublea} that within the standard
approximations, where one includes the bare quark loop as well as the
leading nonperturbative corrections proportional to the quark
condensate and the mixed quark-gluon condensate, there is no
contribution to this function. The leading terms, then, are of order
$\alpha_s$ and come from the two-loop radiative corrections to the
quark loop, the one-loop radiative corrections to the quark condensate,
and the gluon condensate. In general, the calculation of all diagrams
arising at order $\alpha_s$ in the sum rule for a three-current
correlator is rather tedious. In fact, the two-loop corrections to the
triangle diagram have never been computed before. Because of the
particular structure of the trace associated with $\Xi_2$ in
(\ref{correl}), however, the calculation is considerably simplified in
this case. Only the three diagrams shown in Fig.~\ref{fig:1}
contribute. Note that for each contribution the dependence of the
spectral density $\widetilde{\rho}_2^{\rm\,th}(\omega_+,y)$ on
$\omega_+$ is known on dimensional grounds, {\it i.e.}
\begin{eqnarray}\label{wdep}
   \widetilde{\rho}_2^{\rm\,pert} &\,\propto\,& \omega_+^3 ,
    \nonumber\\
   \widetilde{\rho}_2^{\,\langle\bar q q\rangle} &\,\propto\,&
    {\rm const.} , \nonumber\\
   \widetilde{\rho}_2^{\,\langle GG\rangle} &\,\propto\,&
    \delta(\omega_+) .
\end{eqnarray}
Instead of the spectral densities, it thus suffices to calculate
directly the Borel transform of the individual contributions to
$\Xi_2^{\rm th}$, corresponding to the limit $\omega_0\to\infty$ in
(\ref{sumrul}). The $\omega_0$-dependence can then be deduced from
(\ref{wdep}).

Let us now present the results of our calculation. The contributions
from condensates are obtained by evaluating the diagrams shown in
Fig.~\ref{fig:1}(b) and (c), using the one-loop tensor integrals given
in Ref.~\cite{Sublea}. We find
\begin{equation}\label{Xicond}
   \widehat{\Xi}_2^{\rm cond}(T,y) =
   {\alpha_s\langle\bar q q\rangle\,T\over 6\pi}\,
   \bigg[ {1-r(y)\over y-1} + {1\over y+1} \bigg]
   - {\langle\alpha_s GG\rangle\over 96\pi} \bigg({2\over y+1}\bigg) ,
\end{equation}
where the function
\begin{equation}
   r(y) = {1\over\sqrt{y^2-1}}\,\ln\big(y + \sqrt{y^2-1}\big)
\end{equation}
satisfies $r(1)=1$ and $r'(1)=-\case{1}/{3}$, such that there is no
singularity in (\ref{Xicond}) as $y\to 1$. The calculation of the
two-loop diagram depicted in Fig.~\ref{fig:1}(a) is more tedious. In an
intermediate step, one encounters the two-loop tensor integral
\begin{equation}
   \int{{\rm d}^D\!s\over(2\pi)^D}{{\rm d}^Dt\over(2\pi)^D}\,
   {(s-t)_\alpha\,s_\beta\,t_\gamma\over
    (\omega+2 v\!\cdot\!s) (\omega+2 v\!\cdot\!t)
    (\omega'+2 v'\!\cdot\!t)\,s^2\,t^2\,(s-t)^2} .
\end{equation}
For its evaluation it is convenient to represent the light quark
propagators by Fourier transforms in a $D$-dimensional Euclidean space
(see \cite{PT}) and use an exponential integral representation of the
heavy quark propagators. After performing the integrations over the
loop momenta, one is left with five parameter integrals, two of which
become trivial after Borel transformation. The remaining three
integrals stay finite for $D=4$ and can be calculated in closed form.
Our result is
\begin{equation}\label{Xipert}
   \widehat{\Xi}_2^{\rm pert}(T,y) =
   - {\alpha_s T^4\over 8\pi^3}\,\bigg({2\over y+1}\bigg)^2
   \bigg[ {1-r(y)\over y-1} + 2 \bigg] .
\end{equation}
Based on (\ref{sumrul}) and (\ref{wdep}), we can now introduce back the
dependence of the various contributions on the continuum threshold
$\omega_0$ to obtain the final sum rule
\begin{eqnarray}\label{chi2sr}
   \chi_2(y)\,F^2\,\bar\Lambda\,e^{-2\bar\Lambda/T}
   &=& - {\alpha_s T^4\over 8\pi^3}\,\bigg({2\over y+1}\bigg)^2
    \bigg[ {1-r(y)\over y-1} + 2 \bigg]\,
    \delta_3\Big({\omega_0\over T}\Big) \nonumber\\
   &&+ {\alpha_s\langle\bar q q\rangle\,T\over 6\pi}\,
    \bigg[ {1-r(y)\over y-1} + {1\over y+1} \bigg]\,
    \delta_0\Big({\omega_0\over T}\Big) \nonumber\\
   &&- {\langle\alpha_s GG\rangle\over 96\pi}
    \bigg({2\over y+1}\bigg) ,
\end{eqnarray}
where
\begin{equation}
   \delta_n(x) = {1\over\Gamma(n+1)}\int\limits_0^x\!{\rm d}z\,
   z^n\,e^{-z} .
\end{equation}

\section{Evaluation of the Sum Rule}

For the QCD parameters in (\ref{chi2sr}) we take the standard values
\begin{eqnarray}
   \alpha_s/\pi &=& 0.1 , \nonumber\\
   \langle\bar q q\rangle &=& - (0.23\,{\rm GeV})^3 , \nonumber\\
   \langle\alpha_s GG\rangle &=& 0.038\,{\rm GeV^4} .
\end{eqnarray}
The value of the strong coupling corresponds to the scale
$\mu=2\bar\Lambda\simeq 1$ GeV, which is appropriate for evaluating
radiative corrections in the effective theory \cite{SR}. In addition,
the sum rule depends on the continuum threshold $\omega_0$ and on the
Borel parameter $T$. These quantities have recently been determined
from the analysis of a QCD sum rule for the correlator of two
heavy-light currents in the effective theory \cite{SR2}. One finds good
stability for $\omega_0=2.0\pm 0.3\,{\rm GeV}$, and the consistency of
the theoretical calculation requires that the Borel parameter be in the
range $0.6<T<1.0$ GeV.

With this set of parameters one finds $F=0.30\pm 0.05\,{\rm GeV^{3/2}}$
and $\bar\Lambda=0.50\pm 0.07\,{\rm GeV}$ for the hadronic parameters
appearing on the left-hand side of (\ref{chi2sr}). For reasons of
consistency, however, one should rather work directly with the QCD sum
rule for these quantities. This procedure minimizes the $T$-dependence
of the final result. We thus use \cite{SR2}
\begin{equation}\label{Fsum}
   F^2\,\bar\Lambda\,e^{-2\bar\Lambda/T}
   = {9 T^4\over 8\pi^2}\,\delta_3\Big({\omega_0\over T}\Big)
   - {m_0^2\langle\bar q q\rangle\over 4 T} ,
\end{equation}
where $m_0^2\simeq 0.8\,{\rm GeV^2}$ is the ratio of the mixed
quark-gluon condensate and the quark condensate. Combining
(\ref{chi2sr}) and (\ref{Fsum}), we express $\chi_2(y)$ as a function
of $\omega_0$, $T$, and the QCD parameters. The numerical analysis of
the resulting sum rule is shown in Fig.~\ref{fig:2}. We observe
excellent stability against variation of the Borel parameter $T$, and
it supports the self-consistency of the approach that the stability
region coincides with that determined in Ref.~\cite{SR2}. Also the
dependence on the threshold parameter is weak. Varying $\omega_0$
between 1.7 and 2.3 GeV changes $\chi_2(y)$ by no more than $\pm
0.4\%$. For instance, we obtain from Fig.~\ref{fig:2}b the zero recoil
prediction $\chi_2(1)=-(3.8\pm 0.4)\%$.

Finally, we note that a rough estimate of the form factor can be
obtained by neglecting the nonperturbative terms in (\ref{chi2sr}) and
(\ref{Fsum}). Independently of $\omega_0$ and $T$, this gives the
simple function
\begin{equation}
   \chi_2^{\rm pert}(y) =
   - {\alpha_s\over 9\pi}\,\bigg({2\over y+1}\bigg)^2
   \bigg[ {1-r(y)\over y-1} + 2 \bigg] ,
\end{equation}
which is smaller than the complete sum rule result by roughly a factor
$\case{2}/{3}$. At zero recoil, for instance, one finds
$\chi_2^{\rm pert}(1) = -\case{7}/{27} (\alpha_s/\pi)\simeq -2.6\%$.

In summary, we have presented the complete order-$\alpha_s$ QCD sum
rule analysis of the subleading Isgur-Wise function $\chi_2(y)$. We
find that this form factor is small, typically of the order of $-2\%$
to $-4\%$. When combined with the analysis of Ref.~\cite{Sublea}, which
predicts similarly small values for $\chi_3(y)$, but much larger values
for the remaining two subleading form factors $\chi_1(y)$ and
$\xi_3(y)$, these results strongly indicate that power corrections in
the heavy quark expansion which are related to the chromo-magnetic
moment operator are small.

\newpage
\acknowledgements
It is a pleasure to thank A. Radyushkin and A. Falk for helpful
discussions. M.N. gratefully acknowledges financial support from the
BASF Aktiengesellschaft and from the German National Scholarship
Foundation. Y.N. is an incumbent of the Ruth E. Recu Career Development
chair, and is supported in part by the Israel Commission for Basic
Research and by the Minerva Foundation. This work was also supported by
the Department of Energy, contract DE-AC03-76SF00515.

\figure{\label{fig:1}
Diagrams contributing at order $\alpha_s$ to the sum rule for the
universal form factor $\chi_2(v\cdot v')$: (a) perturbative
contribution, (b) quark condensate contribution, (c) gluon condensate
contribution. Heavy quark propagators are drawn as double lines, and
a square represents the chromo-magnetic moment operator.}

\figure{\label{fig:2}
Numerical evaluation of the sum rules (\ref{chi2sr}) and (\ref{Fsum}):
(a) form factor $\chi_2(v\cdot v')$ evaluated for $\omega_0=2\,
{\rm GeV}$ and $0.6<T<1.0\,{\rm GeV}$; (b) dependence of $\chi_2(1)$ on
the Borel parameter $T$ for $\omega_0=2.3$, 2.0, 1.7 GeV (top to
bottom).}

\end{document}